\documentclass[american,aps,manuscript,preprintnumbers]{revtex4-1}
\usepackage[T1]{fontenc}
\usepackage[latin9]{inputenc}
\usepackage{bm}
\usepackage{graphicx}

\makeatletter

\providecommand{\tabularnewline}{\\}

\usepackage{amsmath}
\allowdisplaybreaks

\makeatother

\usepackage{babel}
\begin{document}

\title{Slip velocity of lattice Boltzmann simulation \\
using bounce-back boundary scheme}

\author{Jianping Meng}

\email[To whom correspondence should be addressed: ]{jianping.meng@stfc.ac.uk}

\author{Xiao-Jun Gu}

\email{xiaojun.gu@stfc.ac.uk}

\author{David R Emerson}

\email{david.emerson@stfc.ac.uk}

\affiliation{Scientific Computing Department, STFC Daresbury laboratory, Warrington
WA4 4AD, United Kingdom}

\date{\today}
\begin{abstract}
In this work we investigate the issue of non-physical slip at wall
of lattice Boltzmann simulations with the bounce-back boundary scheme.
By comparing the analytical solution of two lattice models with four
and nine discrete velocities for the force-driven Poiseuille flow,
we are able to reveal the exact mechanism causing the issue. In fact,
no boundary condition is defined by the bounce-back scheme for the
the discrete velocities parallel to wall. Other factors, such as initial
conditions and inlet and outlet boundary conditions, can play the
role and induce the non-physical slip velocity. Therefore, the issue
is not related to the single-relaxation-time scheme. Naturally the
key for resolving it is to specify the definition for these velocities.
Through a lid-driven cavity flow, we show that the solution can be
as easy as no extra effort required for simple geometries, although
further study is necessary for complex geometries. 
\end{abstract}



\maketitle

\section{Introduction\label{sec:Section-I}}

The lattice Boltzmann method (LBM) has been developed as a mesocopic
computational fluid dynamics (CFD) tool for the Navier-Stokes (NS)
level problems and beyond \citep{doi:10.1146/annurev.fluid.30.1.329,Qian1995,Aidun2010}.
Due to its origin from the lattice gas automata (LGA) \citep{PhysRevLett.56.1505},
it keeps the flexibility of a particle method to a great extent. On
the other hand,the stochastic noise is eliminated in LBM by using
the distribution function. Importantly, this links the LBM into the
discrete velocity method of the Boltzmann-BGK (Bhatnagar-Gross-Krook)
equation\citep{PhysRevE.55.R6333,PhysRevE.56.6811,PhysRevLett.80.65,2006JFM550413S}.
From such a point of view, we actually solve a system of partial differential
equations with linear advection terms. This opens the door of introducing
more sophisticated scheme leading to such as finite difference LBM
or finite volume LBM (e.g., \citep{Mei1998} and \citep{PhysRevE.59.6202}). 

It is the simplicity which brings the popularity of LBM. The algorithm
is easy to understand for application purpose. Since an explicit scheme
is employed, the programming and parallelism is straightforward. The
second order accuracy in both space and time is achieved at the expanse
of a first order scheme, which is fairly enough for most purposes.
The boundary treatment, even for complex geometry, can be incredibly
simple due to the so-called bounce-back (BB) scheme which only requires
the particles to reverse their velocity on the wall/obstacle \citep{Cornubert1991241,He1997}
.

However, there are non-physical slip velocities occurring at wall
in simulations using the BB scheme. This was firstly discovered for
two LGA models \citep{Cornubert1991241} and then was analysed for
the nine-discrete-velocity (D2Q9) LBM in \citep{He1997}. By using
a simple force-driven Poiseuille flow, it was shown that the non-physical
slip can be generated on the wall with the BB family scheme \citep{He1997}.
Since the slip velocity was found to be related to the mesh size,
it was then deemed as a numerical artificial effect. Later, this issue
has been considered as an inherent deficiency of the single-relaxation-time
(SRT) scheme since it may be resolved by using extra free parameters
in two-relaxation-time (TRT) or multi-relaxation time (MRT) schemes
(see e.g. \citep{PhysRevE.83.056710}). Indeed, it is believed that
the SRT plus BB combination cannot avoid this issue \citep{Pan2006898}.

However, due to its simplicity, the SRT plus BB combination is more
favourable for application purpose. Therefore, it is of interest to
investigate how the slip velocity is induced and therefore gain useful
information on how to fix it. For this purpose, we will first analysis
a lattice model with four discrete velocities (D2Q4) for the force-driven
Poiseuille flow following the method presented in \citep{He1997}.
With this lattice model, we will see that the SRT plus BB combination
does not necessarily induce non-physical slip velocity and it is possible
to correctly implement the non-slip wall. Then we will compare this
model with the D2Q9 model to find the exact mechanism inducing the
slip velocity. With these findings, we devise a guidance on how to
fix the non-physical slip velocity. Finally, we will examine this
guidance and thereby the discussions on the mechanism by simulating
the lid-driven cavity flow with the D2Q9 lattice.

\section{Lattice Boltzmann scheme and lattices}

The LBM can be considered as an approximation to the Boltzmann-BGK
equation \citep{PhysRevE.55.R6333,PhysRevE.56.6811,PhysRevLett.80.65,2006JFM550413S}.
After the discretisation in the particle velocity space, the governing
equation becomes 
\begin{equation}
\frac{\partial f_{\alpha}}{\partial t}+\bm{C}_{\alpha}\cdot\nabla f_{\alpha}=-\frac{1}{\tau}(f_{\alpha}-f_{\alpha}^{eq})+F_{\alpha},\label{pde}
\end{equation}
which represents the evolution of the distribution function $f_{\alpha}(\bm{r},t)$
for the $\alpha$th discrete velocity $\bm{C}_{\alpha}$ at position
$\bm{r}=(x,y,z)$ and time $t$. The effect of external body force
is described by $F_{\alpha}$. The particle interaction is modelled
by a relaxation term towards the discrete equilibrium distribution
function $f_{\alpha}^{eq}(\bm{r},t)$. In order to simulate incompressible
and isothermal flows, it is common to use an equilibrium function
with second order velocity terms, i.e., 
\begin{equation}
f_{\alpha}^{eq}=w_{\alpha}\rho[1+\frac{\bm{U}\cdot\bm{C}_{\alpha}}{RT_{0}}+\frac{1}{2}\frac{(\bm{U}\cdot\bm{C}_{\alpha})^{2}}{(RT_{0})^{2}}-\frac{\bm{U}\cdot\bm{U}}{2RT_{0}}],\label{feq}
\end{equation}
which is determined by the density, $\rho$, the fluid velocity, $\bm{U}$,
and the reference temperature, $T_{0}$. For gas flows, the constant,
$R$, can be conveniently understood as the gas constant. If a liquid
fluid is involved, it, together with $T_{0}$, can be considered as
a reference quantity. For convenience, the sound speed $c_{s}$ is
often considered equal to $\sqrt{RT_{0}}$, although there is a constant
factor of difference. The weight factor is denoted by $w_{\alpha}$
for the discrete velocity $\bm{C}_{\alpha}$. The term \$F\_\textbackslash{}alpha\$
can be obtained by using various method \citep{2006JFM550413S,He1998a}.
Here, the first order expansion is sufficient for our purpose, which
can be written as \citep{2006JFM550413S}.
\begin{equation}
F_{\alpha}=\rho w_{\alpha}\frac{\bm{G}\cdotp\bm{C}_{\alpha}}{RT_{0}}.
\end{equation}
where the actual induced acceleration is denoted by $\bm{G}$. The
relaxation time , $\tau$, is related to the fluid viscosity , $\mu$,
and the pressure, $p$, via the Chapman-Enskog expansion, i.e., $\mu=p\tau$.
Hence, for isothermal and incompressible flows, the Reynolds number
becomes $Re=\rho_{0}U_{0}L/\mu=U_{0}L/(\tau RT_{0})$, where we use
a subscript $0$ to denote the reference value and $L$ the characteristic
length of the system. It is worth noting here that the Knudsen number
can be defined as $\mu_{0}\sqrt{RT_{0}}/(p_{0}L)$. So, the relaxation
time $\tau$ is also related to Knudsen number by the viscosity $Kn=\tau\sqrt{RT_{0}}/L,$
where $p_{0}=\rho_{0}RT_{0}$ is applied. In this sense, we have $Kn\times Re=U_{0}/\sqrt{RT_{0}}=Ma$.
To get the density and velocity, we only need summation operations,
i.e., 
\[
\:\rho=\sum_{\alpha}f_{\alpha},\;\mbox{{and},\;}\rho\bm{U}=\sum_{\alpha}f_{\alpha}\bm{C}_{\alpha}.
\]

To numerically solve Eq.~(\ref{pde}), a smart trapezoidal scheme
can used to achieve the particle-jump like simulation \citep{He1998},
which can be written as

\begin{equation}
\tilde{f}_{\alpha}(\bm{r}+\bm{C}_{\alpha}dt,t+dt)-\tilde{f}_{\alpha}(\bm{r},t)=-\frac{dt}{\tau+0.5dt}\left[\tilde{f}_{\alpha}(\bm{r},t)-f_{\alpha}^{eq}(\bm{\bm{r}},t)\right]+\frac{\tau F_{\alpha}dt}{\tau+0.5dt},\label{eq:scheme}
\end{equation}
where
\begin{equation}
\tilde{f}_{\alpha}=f_{\alpha}+\frac{dt}{2\tau}(f_{\alpha}-f_{\alpha}^{eq})-\frac{dt}{2}F_{\alpha}.
\end{equation}
By using $\tilde{f}_{\alpha}$ this scheme is ready for implementing
the stream-collision algorithm. At the same time, the macroscopic
quantities become

\begin{equation}
\rho=\sum_{\alpha}\tilde{f}_{\alpha},\ \mbox{{and},\ }\rho\bm{U}=\sum_{\alpha}\bm{C}_{\alpha}\tilde{f}+\frac{\rho\bm{G}dt}{2}.\label{eq:hemac}
\end{equation}

For two dimensional flows, the D2Q9 lattice is commonly used where
the nine discrete velocities ($\alpha=1..9$) are 
\begin{equation}
C_{\alpha,x}=\sqrt{3RT_{0}}[0,1,0,-1,0,1,-1,-1,1],
\end{equation}
\begin{equation}
C_{\alpha,y}=\sqrt{3RT_{0}}[0,0,1,0,-1,1,1,-1,-1],
\end{equation}
and the corresponding weights are
\begin{equation}
w_{\alpha}=[\frac{4}{9},\frac{1}{9},\frac{1}{9},\frac{1}{9},\frac{1}{9},\frac{1}{36},\frac{1}{36},\frac{1}{36},\frac{1}{36}].
\end{equation}

As discussed above, the stream-collision algorithm is ready to be
implemented now. The only trick is to tie the space and time step
together as $d\bm{r}=\bm{C}_{\alpha}dt$. For instance, assuming the
system length is $L$, we may set the spatial step $dx=L/N$ and then
$dt=L/(N\sqrt{3RT_{0}})$ where $N$ is the cell number. This insures
the ``particles'' are jumping on a uniform grid system. In simulations,
it is common practice to use a non-dimensional system in which the
space and time step are considered as reference value. Apparently,
this will make no difference on results. However, confusion may be
caused in this way. We shall return to this point below. Alternatively,
we may also transform Eq.~(\ref{pde}) to its non-dimensional form
first by using the reference values presented in \citep{Meng2011b}
and then apply the scheme Eq.~(\ref{eq:scheme}). Again, this non-dimensional
transformation will not alter the final simulation results but the
relations with dimensional quantities are more clear, at least for
gas dynamics. 

\begin{figure}
\includegraphics[scale=0.5]{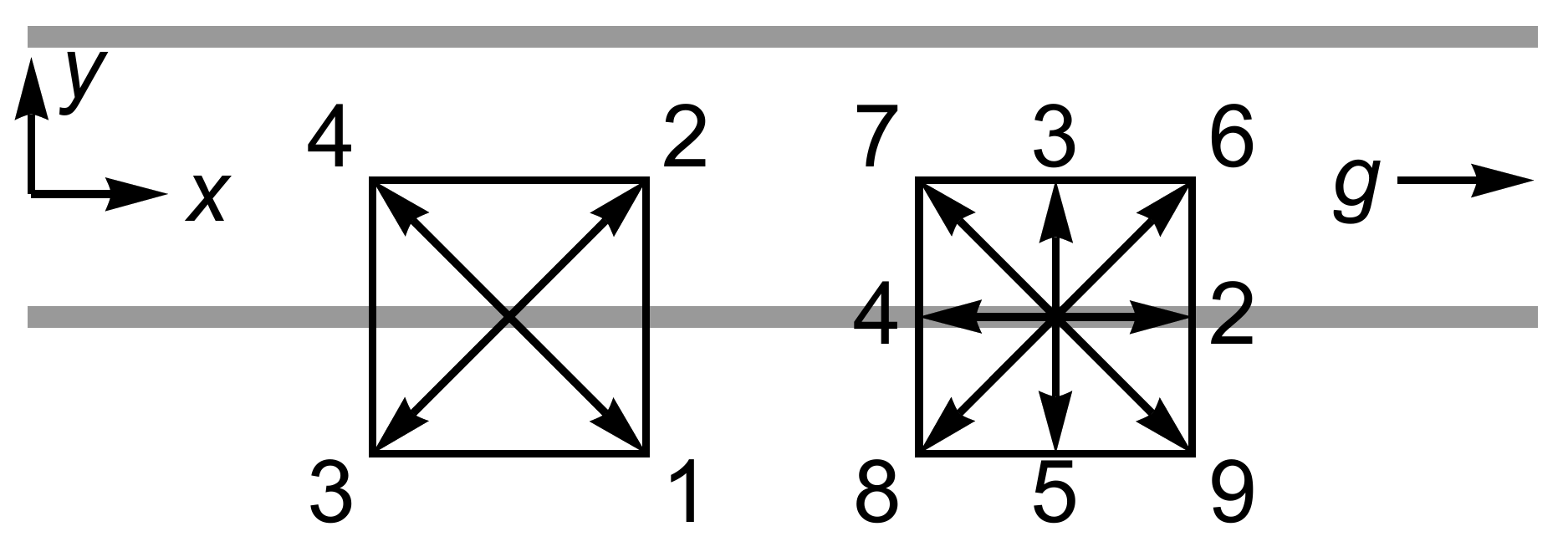}

\protect\caption{Illustration of the D2Q4 and D2Q9 lattice at the bottom wall for a
force-driven Poiseuille flow. The discrete velocity $\bm{C}_{1}=\left(0,0\right)$
of the D2Q9 lattice is not shown.}
\label{Illu}
\end{figure}

To study the effect of the BB scheme on the solid boundary, we will
first use a D2Q4 model \citep{2006JFM550413S} where the four discrete
velocities are 
\begin{equation}
C_{\alpha,x}=\sqrt{RT_{0}}[1,1,-1,-1],
\end{equation}
\begin{equation}
C_{\alpha,y}=\sqrt{RT_{0}}[-1,1,-1,1],
\end{equation}
and the weights are

\begin{equation}
w_{\alpha}=[\frac{1}{4},\frac{1}{4},\frac{1}{4},\frac{1}{4}].
\end{equation}
Both the D2Q9 and D2Q4 models are illustrated in Fig. \ref{Illu}.
It is worth noting that in the D2Q4 model there is no discrete velocity
parallel to the wall for the force-driven Poiseuille flow with regular
shaped channel (i.e., the wall is either horizontal or vertical),
which will make a dramatical difference from the D2Q9 model for the
force-driven Poiseuille flow.

\section{Slip velocity and bounce-back scheme}

\subsection{Solution of D2Q4 model for force-driven Poiseuille flow}

In the following, we will use the method presented in \citep{He1997}
to solve the D2Q4 model for the force-driven Poiseuille flow. For
convenience, we introduce some new notations $\bm{G/}\sqrt{RT_{0}}=(g_{x},g_{y})$,
$\bm{U/}\sqrt{RT_{0}}=(u,v)$ and $\bm{C/}\sqrt{RT_{0}}=\bm{c}=(c_{x},c_{y})$.
This is just for the simplicity of formulations and should not be
understood as a non-dimensional transformation in this work. Instead,
we stick to the dimensional system presented in Eqs. (\ref{pde})
and (\ref{eq:scheme}). Moreover, for convenience, we use letter $A$
to denotes the coefficient $dt/(\tau+0.5dt)\frac{}{}$ of the relaxation
term and $B$ for the coefficient $\tau dt/(\tau+0.5dt)$ of the force
term. Without influencing the discussion, we set $g_{y}$ and $v$
to be zero. Therefore, for the D2Q4 lattice, the evolutionary rules
for the $j$th bulk node are 

\begin{subequations} \label{eq:d2q4r1}
\begin{align}
\tilde{f}_{j,1} &=(1-A)\tilde{f}{}_{j+1,1}+\frac{1}{4}A\rho u_{j+1}+\frac{A\rho}{4}+\frac{Bg_{x}\rho}{4}, \\
\tilde{f}_{j,2} &=(1-A)\tilde{f}_{j-1,2}+\frac{1}{4}A\rho u_{j-1}+\frac{A\rho}{4}+\frac{Bg_{x}\rho}{4}, \\
\tilde{f}_{j,3} &=(1-A)\tilde{f}_{j+1,3}-\frac{1}{4}A\rho u_{j+1}+\frac{A\rho}{4}-\frac{Bg_{x}\rho}{4}, \\
\tilde{f}_{j,4} &=(1-A)\tilde{f}_{j-1,4}-\frac{1}{4}A\rho u_{j-1}+\frac{A\rho}{4}-\frac{Bg_{x}\rho}{4},
\end{align}
\end{subequations}while Eq. (\ref{eq:hemac}) for the velocity becomes 

\begin{equation}
\rho u_{j}=\tilde{f}_{j,1}+\tilde{f}_{j,2}-\tilde{f}_{j,3}-\tilde{f}_{j,4}+\frac{\rho g_{x}dt}{2}.\label{eq:d2q4u}
\end{equation}
To get the macroscopic governing equation, we follow the procedure
of \citep{He1997}. For convenience, we write a few alternative variants
of rules Eqs.(\ref{eq:d2q4r1})-(\ref{eq:d2q4u}), i.e.,

\begin{subequations} \label{eq:d2q4r2}
\begin{align}
\tilde{f}_{j-1,1}&=(1-A)\tilde{f}_{j,1}+\frac{1}{4}A\rho u_{j}+\frac{A\rho}{4}+\frac{Bg_{x}\rho}{4}, \label{eq:d2q4r2a}\\
\tilde{f}_{j+1,2}&=(1-A)\tilde{f}_{j,2}+\frac{1}{4}A\rho u_{j}+\frac{A\rho}{4}+\frac{Bg_x\rho}{4},\\
\tilde{f}_{j-1,3}&=(1-A)\tilde{f}_{j,3}-\frac{1}{4}A\rho u_{j}+\frac{A\rho}{4}-\frac{Bg_x\rho}{4},\label{eq:d2q4r2c} \\ 
\tilde{f}_{j+1,4}&=(1-A)\tilde{f}_{j,4}-\frac{1}{4}A\rho u_{j}+\frac{A\rho}{4}-\frac{Bg_x\rho}{4},
\end{align}
\end{subequations}and

\begin{subequations} \label{eq:d2q4u1}
\begin{align}
\rho u_{j+1}=\tilde{f}_{j+1,1}+\tilde{f}_{j+1,2}-\tilde{f}_{j+1,3}-\tilde{f}_{j+1,4}+\frac{dt g_{x}\rho}{2},\\
\rho u_{j-1}=\tilde{f}_{j-1,1}+\tilde{f}_{j-1,2}-\tilde{f}_{j-1,3}-\tilde{f}_{j-1,4}+\frac{dt g_{x}\rho}{2}.\label{eq:d2q4u1b}
\end{align}
\end{subequations}Applying the rules Eq. (\ref{eq:d2q4r1}) into Eq. (\ref{eq:d2q4u}),
we have
\begin{equation}
\rho u_{j}=(1-A)(\tilde{f}_{j-1,2}-\tilde{f}_{j-1,4}+\tilde{f}_{j+1,1}-\tilde{f}_{j+1,3})+\frac{1}{2}A\rho u_{j-1}+\frac{1}{2}A\rho u_{j+1}+Bg_{x}\rho+\frac{dtg_{x}\rho}{2}.
\end{equation}
Hence, we need to work out $(1-A)(\tilde{f}_{j-1,2}-\tilde{f}_{j-1,4}+\tilde{f}_{j+1,1}-\tilde{f}_{j+1,3})$.
To do so, the main idea is to use the rules Eqs.(\ref{eq:d2q4r1})
and (\ref{eq:d2q4r2}) and Eqs. (\ref{eq:d2q4u}) and (\ref{eq:d2q4u1})
alternatively. The aim is to relate the unknown distribution functions
to the macroscopic quantities. For example, to obtain $\tilde{f}_{j-1,2}-\tilde{f}_{j-1,4}$,
we first use (\ref{eq:d2q4u1b}) to represent the unknowns with $u_{j-1}$,
$\tilde{f}_{j-1,1}$ and $\hat{f}_{j-1,3}$, then use rules (\ref{eq:d2q4r2a})
and (\ref{eq:d2q4r2c}) to transform $\tilde{f}_{j-1,1}-\tilde{f}_{j-1,3}$
into a formula of $u_{j}$, $\tilde{f}_{j,1}$ and $\tilde{f}_{j,3}$.
Finally, we can apply the rule Eq. (\ref{eq:d2q4u}) to convert all
$f_{j}$s into a formula of $u_{j}$. Following this idea and through
a few iterations, we obtain 
\begin{equation}
Ag_{x}\rho(Adt+2B)+(2-A)\rho u_{j-1}+2(A-2)\rho u_{j}+(2-A)\rho u_{j+1}=0.
\end{equation}
Considering the meaning of $A$ and $B$, the equation becomes
\begin{equation}
\rho dt^{2}g_{x}+\rho\tau(u_{j-1}-2u_{j}+u_{j+1})=0.
\end{equation}
Further using $dt=dx/\sqrt{RT_{0}}$ , $\mu=p\tau$,and $p_{0}=\rho_{0}RT_{0}$,
the final form is 
\begin{equation}
\frac{\mu(u_{j-1}-2u_{j}+u_{j+1})}{dx^{2}}+\rho g_{x}=0,\label{eq:lbmns}
\end{equation}
which is exactly the second central difference scheme of the NS equations
for this simple force-driven flow. It has a simple solution

\begin{equation}
u_{j}=\rho g\frac{(L-jdx)jdx}{2\mu}+U_{s},\thinspace\thinspace\thinspace j=0,1,2,3,\cdots N.\label{eq:res}
\end{equation}
The slip velocity is denoted by $U_{s}$, which is produced by the
boundary treatment, either physically or non-physically. So, its exact
value will depend on the specific boundary condition.

To find $U_{s}$, we also follow the procedure of \citep{He1998}.
First, we introduce a notation $U_{0}=\sum_{\alpha}c_{\alpha,x}f_{0,\alpha}$
where the incoming distribution functions will be determined by the
boundary condition. In order to find the slip velocity, we actually
look at the node $j=1$. Following the manner of finding the bulk
equation, we will be able to obtain the relation of the prescribed
boundary speed $u_{0}$, $u_{1}$, $u_{2}$ and $U_{0}$. The trick
is that the prescribed boundary velocity $(u_{0},0)$ is used when
applying Eq. (\ref{eq:d2q4r1}) into Eq. (\ref{eq:d2q4u}). However,
when using the rule (\ref{eq:d2q4u1b}) for $j=1$, we need to consider
the relation $U_{0}=\sum_{\alpha}c_{\alpha,x}f_{0,\alpha}$. Through
simple calculations, the relation can be written as 
\begin{equation}
u_{1}=\frac{u_{0}+u_{2}}{2}+\frac{dt^{2}g_{x}}{2\tau}-\frac{(dt-2\tau)\left(U_{0}-u_{0}\right)}{4\tau}.\label{eq:u1}
\end{equation}
 For simplicity, we assume a non-slip boundary with zero speed at
wall in the following, i.e., $u_{0}$ is set to be zero. Then, substituting
the solution (\ref{eq:res}) into Eq. (\ref{eq:u1}), we can obtain
the slip velocity

\begin{equation}
U_{s}=\frac{2\tau-dt}{2\tau}U_{0}.\label{eq:us}
\end{equation}

For the so called modified BB (MBB) scheme (means that collision and
forcing still occur at boundary nodes \citep{He1997}), the rule at
the boundary point is 

\begin{equation}
f_{0,4}=f_{0,1}\ \ f_{0,2}=f_{0,3}.
\end{equation}
It can be seen that $U_{0}$ must be zero. Hence, for the D2Q4 model,
the slip velocity $U_{s}$ is zero. Although the SRT scheme is used,
the MBB scheme leads to a correct non-slip boundary condition. Moreover,
if rotating the wall direction from horizontal to vertical, it can
be easily seen that slip velocity will also be zero. 

For the BB scheme without collision and forcing occurring at boundary
nodes, we can not directly apply Eqs. (\ref{eq:u1}) and (\ref{eq:us}).
But it is straightforward to see that $U_{s}$ will be zero. 

As has been shown, the results of D2Q4 model are significantly different
from that of the D2Q9 model and D2Q5 model presented in \citep{Cornubert1991241,He1997}.
This helps to clarify the relation between the SRT scheme and non-physical
slip velocity. The SRT scheme plus the MBB or BB boundary scheme does
not necessarily induces non-physical slip velocity.

\subsection{Mechanism of non-physical slip velocity with D2Q9 model. }

Now it is natural to ask why there is non-physical slip velocity in
such as the D2Q9 solution. For this purpose, we return to the D2Q9
model following \citep{He1997}. The rules for the D2Q9 model are

\begin{subequations}\label{eq:d2q9rule}
\begin{align}
\tilde{f}_{j,1}&=\frac{4\rho}{9}-\frac{2\rho u_{j}^{2}}{9}\\
\tilde{f}_{j,2}&=\frac{\rho u_{j}^{2}}{9}+\frac{\rho u_{j}}{3\sqrt{3}}+\frac{Bg_{x}\rho}{3\sqrt{3}A}+\frac{\rho}{9}\label{eq:d2q9rule2}\\ 
\tilde{f}_{j,3}&=-\frac{1}{18}A\rho u_{j-1}^{2}+\frac{A\rho}{9}+(1-A)\tilde{f}_{j-1,3}\\
\tilde{f}_{j,4}&=\frac{\rho u_{j}^{2}}{9}-\frac{\rho u_{j}}{3\sqrt{3}}-\frac{Bg_{x}\rho}{3\sqrt{3}A}+\frac{\rho}{9}\label{eq:d2q9rule4}\\ 
\tilde{f}_{j,5}&=-\frac{1}{18}A\rho u_{j+1}^{2}+\frac{A\rho}{9}+(1-A)\tilde{f}_{j+1,5}\\
\tilde{f}_{j,6}&=\frac{1}{36}A\rho u_{j-1}^{2}+\frac{A\rho u_{j-1}}{12\sqrt{3}}+\frac{A\rho}{36}+\frac{Bg_{x}\rho}{12\sqrt{3}}+(1-A)\tilde{f}_{j-1,6}\\
\tilde{f}_{j,7}&=\frac{1}{36}A\rho u_{j-1}^{2}-\frac{A\rho u_{j-1}}{12\sqrt{3}}+\frac{A\rho}{36}-\frac{Bg_{x}\rho}{12\sqrt{3}}+(1-A)\tilde{f}_{j-1,7}\\
\tilde{f}_{j,8}&=\frac{1}{36}A\rho u_{j+1}^{2}-\frac{A\rho u_{j+1}}{12\sqrt{3}}+\frac{A\rho}{36}-\frac{Bg_{x}\rho}{12\sqrt{3}}+(1-A)\tilde{f}_{j+1,8}\\
\tilde{f}_{j,9}&=\frac{1}{36}A\rho u_{j+1}^{2}+\frac{A\rho u_{j+1}}{12\sqrt{3}}+\frac{A\rho}{36}+\frac{Bg_{x}\rho}{12\sqrt{3}}+(1-A)\tilde{f}_{j+1,9}.
\end{align}
\end{subequations}As discussed before, we will pay particular attention to discrete
velocities parallel to the wall. With the horizontal wall, they are
the 2nd and 4th velocity as shown in Fig. \ref{Illu}. Looking at
Eqs. (\ref{eq:d2q9rule2}) and (\ref{eq:d2q9rule4}), we remind that
they are mainly the consequence of periodic boundary conditions for
the inlet and outlet, i.e., there is no gradient in the streamwise
direction. In other words, they are not solely related to the SRT
scheme and are NOT determined by the BB scheme at all. Similarly,
the governing equation for bulk nodes is

\begin{equation}
\frac{\mu(u_{j-1}-2u_{j}+u_{j+1})}{dx^{2}}+3dt_{9}^{2}\rho g_{x}=0,
\end{equation}
where $dt_{9}$ means the time step for the D2Q9 model. Assuming the
space step is same for both two models, $dt_{9}=dt/\sqrt{3}$ where
$dt$ is time step for the D2Q4 model.  Hence two models yield same
governing equation for momentum. For brevity, we only discuss the
MBB scheme. Therefore,$u_{1}$ and $U_{s}$ can be written as

\begin{equation}
u_{1}=\frac{u_{0}+u_{2}}{2}+\frac{dt_{9}^{2}g_{x}}{2\tau}-\frac{3(dt_{9}-2\tau)\left(U_{0}-u_{0}\right)}{4\tau}
\end{equation}
and

\begin{equation}
U_{s}=\frac{3(2\tau-dt_{9})}{2\tau}U_{0}.
\end{equation}
It can be seen that the form of $U_{s}$ is consistent with Eq. (18)
in \citep{He1997}. To find $U_{0}$ we need to calculate out 
\begin{equation}
f_{0,2}-f_{0,4}+f_{0,6}-f_{0,7}+f_{0,9}-f_{0,8}.
\end{equation}
Applying the MBB rule 

\begin{equation}
f_{0,7}=f_{0,9}\ \ f_{0,6}=f_{0,8}\ \ f_{0,3}=f_{0,5},
\end{equation}
we only need to consider $f_{0,2}-f_{0,4}$. After simple calculations
using Eqs. (\ref{eq:d2q9rule2}) and (\ref{eq:d2q9rule4}) (note that
the MBB rule allows collisions at boundary, and again, these two equations
are actually determined by the periodic boundary condition) with the
prescribed boundary velocity $(u_{0}=0,0)$, we find it equals 
\begin{equation}
f_{0,2}-f_{0,4}=\frac{2}{3}g_{x}\rho\tau.
\end{equation}
Regarding that 
\begin{equation}
g_{x}=\frac{8\mu u_{m}}{L^{2}\rho},
\end{equation}
where $u_{m}$ denotes the centerline speed without slip velocity
at boundaries, $U_{s}$ is written as 

\begin{equation}
U_{s}=-g_{x}(dt_{9}-2\tau)=-\frac{8\mu(dt_{9}-2\tau)u_{m}}{L^{2}\rho},\label{eq:uspform}
\end{equation}
which is in the form of physical dimension. To transform to the commonly
used lattice unit, we uses relations 

\begin{equation}
\mu=\rho RT_{0}\tau\ \ L=Ndx=Ndt_{9}\sqrt{3RT_{0}}
\end{equation}
and
\begin{equation}
\hat{\tau}=\frac{\tau}{dt_{9}}+\frac{1}{2},
\end{equation}
which yields
\begin{equation}
U_{s}=\frac{8(\hat{\tau}-1)(2\hat{\tau}-1)u_{m}}{3N^{2}}.\label{eq:uslform}
\end{equation}
Here the units of $U_{s}$ and $u_{m}$ is not important since they
cancel each other. We note the form Eq. (\ref{eq:uslform}) is slightly
different from Eq. (22) in \citep{He1997}. This is mainly because
of difference of the factor $B$, i.e., the treatment of the body
force term, cf. Eq. (\ref{eq:scheme}) and Eq. (1) in \citep{He1997}. 

Clearly, the non-physical slip velocity obtained in \citep{He1997}
is due to the contribution of $f_{0,2}$ and $f_{0,4}$. However,
as we have stressed, they are mainly the consequence of periodic boundaries
at the streamwise direction. Therefore, the failure of the MBB scheme
is due to lack of definition on the behaviour of discrete velocities
parallel to wall. Then, they are actually controlled by other factors.
In this case, it is the boundary scheme used in the inlet and outlet,
which is not the bounce-back scheme. 

In this way, the slip velocity may be arbitrary in numerical practice,
which may depend on the specific inlet and outlet boundary condition,
geometry, other numerical operations at the boundary, and even the
initial condition at wall. 

By identifying the mechanism of non-physical slip velocity, we may
be able to devise remedy for the BB scheme. The key is to supplement
the definition for the behaviour of the discrete velocities parallel
to wall if there are any. Since other distribution function pairs
can cancel each other when obtaining the velocity, they must also
be able to cancel each other so that the velocity is zero. For instance,
in this force-driven Poiseuille flow, although the bulk points must
admit the consequence of the inlet and outlet boundary conditions,
wall boundary points do not have to do so. In other words, we do not
necessarily need to apply rules Eqs. (\ref{eq:d2q9rule2}) and (\ref{eq:d2q9rule4})
which has been done above and in \citep{He1997}. By contrary, We
may initially set $f_{0,2}$ and $f_{0,4}$ to be a equilibrium distribution
with zero velocity and they can remain their initial state all the
time. This simple fix is able to correctly yield zero slip velocity. 

In practice, this can be incredibly easy for simple geometries. In
the following section, we will show that actually no extra effort
is necessary for a lid-driven cavity flow. However, the solution for
complex geometries may need further investigation. 

On the other hand, it is worth noting that extra care may be necessary
when using Eq. (\ref{eq:uslform}) to analyse the accuracy. At a first
glance, $U_{s}$ seems to be a second order small quantity. However,
in Eq.(\ref{eq:uspform}), it is actually more or less a constant
error since we should set $\ensuremath{dt_{9}}$ to be smaller than
$\ensuremath{\tau}$ for stability while both $g_{x}$ and $\tau$
are constant for a given incompressible and isothermal flow configuration.
In our view, that is the confusion caused by using numerical time/spatial
steps as reference quantities.

\subsection{D2Q9 simulations for lid-driven cavity flow}

In this section, we will show how to utilise the above observation
to devise remedy for the non-physical slip velocity. For this purpose,
we will simulate the lid-driven cavity flow using the D2Q9 model.
At the bottom, left and right wall, we will implement the bounce-back
scheme. The non-equilibrium bounce-back scheme \citep{zou:1591} is
adopted for the top moving wall to bring in a wall velocity. 

For the cavity flow, we notice a fact that, for the discrete velocities
parallel to wall, their distributions at boundary nodes are never
affected by those of bulk. For the BB scheme, they will only be affected
by their neighbours at wall. For the MBB scheme, the local collisions
will also come into play. This fact can be utilised for eliminating
the slip velocity. 

For the BB scheme, it can be easily seen that, for the discrete velocities
parallel to wall (e.g., $\bm{C}_{2}$ and $\bm{C}_{4}$ for the bottom
wall), the initial state is actually maintained in a way that the
information is cycling among wall nodes. If the initial conditions
at all wall nodes are set to be the uniform equilibrium distribution
with zero velocity, the distribution of such as $\bm{C}_{2}$ and
$\bm{C}_{4}$ will always be able to cancel each other when finding
the velocity. Therefore, the non-slip velocity boundary can be achieved
without any extra effort. The corner points at the top wall are a
little more tricky as they are singular points. In practice, they
may be treated as either a top wall point or a point of the left or
right wall. Here, to maintain the benefit of ``no extra'' effort,
we need to treat them as a left or right wall point. Otherwise, distributions
at the left and right wall will be affected by those of the top wall
which are changing with time, and the initial equilibrium state with
zero velocity will break down. The other two corner points can treated
in normal way although more discrete velocities need to be ``bounced
back''. 

For the MBB scheme, as collisions will occur at boundary points, the
initial state cannot be maintained. However, using the fact that the
information can not propagate into the bulk for the discrete velocities
parallel to the wall, we are able to blend the relevant distribution
to obtain the nonslip condition at wall. For instance, for the bottom
wall, we can use the average of distribution of $\bm{C}_{2}$ and
$\bm{C}_{4}$ as their new value after the streaming step. 

Numerical simulations are conducted for both two ways with four different
Reynolds numbers while the top wall speed is fixed as $0.1\sqrt{RT_{0}}$.
To examine the speed at the bottom, left, and right wall, we calculate
the sum of $\sqrt{\bm{U}\cdot\bm{U}}$ of all nodes at these three
walls at every time step and then obtain the average speed per time
step and per node. Since we are not examine the solution accuracy,
no convergence test will be done for mesh size. By contrary, we will
use as coarse mesh as possible to obtain results quickly. The maximum
time step is set to be $10,0000$ iterations. While it is not of interest
if the steady state is approached, the first order time derivative
of the $L^{2}$ norm error of velocity is found to be smaller than
$3.5\times10^{-4}$ except for cases with $Re=1000$. The results
are summarised in Table \ref{TabCavity}. As has been shown, the speed
at walls are effectively zero within the machine resolution (double
precision). It is worth noting again that actually no extra effort
is necessary for the BB scheme. 

\begin{table}
\begin{tabular}{|c|c|c|c|c|}
\hline 
 & $Re=10$ & $Re=100$ & $Re=500$ & $Re=1000$\tabularnewline
\hline 
\hline 
Average speed (no collision at wall,$\times10^{-18}$)  & 8.774 & 8.740 & 8.831 & 8.844\tabularnewline
\hline 
Average speed (with collision at wall,$\times10^{-18}$)  & 11.663 & 11.594 & 11.596 & 11.610\tabularnewline
\hline 
\end{tabular}

\protect\caption{Average speed at the bottom, left, and right wall. The top wall speed
is $0.1\sqrt{RT_{0}}$.}
\label{TabCavity}

\end{table}

\section{Concluding remarks}

To conclude, we have investigated the issue of the slip velocity at
wall boundaries in lattice Boltzmann simulations with the BB scheme
family. To identify the mechanism, we have analytically compared the
solutions of a D2Q4 lattice and the commonly used D2Q9 lattice for
the force-driven Poiseuille flow. It is found that the BB family scheme
does not define the behaviour of discrete velocities parallel to the
wall. Mathematically the boundary condition is \textbf{incompletely
determined}. This gives opportunities for other factors to affect
them, such as the boundary conditions for inlet and outlet or even
the initial condition at boundary. The non-physical slip velocity
are induced exactly by these undesired effects. Therefore, the scheme
for the bulk (e.g., the SRT scheme) is not the intrinsic reason for
the slip velocity. 

To solve this issue, the key is to supplement the definition for the
discrete velocities parallel to wall. By simulating the lid-driven
cavity flows, We have shown that this can be incredibly easy for simple
geometries. In fact, we may need no extra effort . The future study
is to find if there is similar solution for complex geometries, which
is already under progress. 
\begin{acknowledgments}
The authors would like to thank the Engineering and Physical Science
Research Council (EPSRC) for their support of Collaborative Computational
Project 12 and 5. Jianping Meng would like to thank Prof. Chao-an
Lin at the National Tsing Hua University for his help on the procedure
presented in \citep{He1997}. 
\end{acknowledgments}

%

\end{document}